# Smart Blockchain Networks: Revolutionizing Donation Tracking in the Web 3.0

Chaimaa Nairi[1], Murtaza Cicioğlu[1*], Ali Çalhan[2]

[1]Department of Computer Engineering, Bursa Uludağ University, 16059, Bursa, Türkiye
[2]Department of Computer Engineering, Düzce University, 81620, Düzce, Türkiye
*Corresponding author: murtazacicioglu@uludag.edu.tr

*Abstract*

A donation-tracking system leveraging smart contracts and blockchain technology holds transformative potential for reshaping the landscape of charitable giving, especially within the context of Web 3.0. This paper explores how smart contracts and blockchain can be used to create a transparent and secure ledger for tracking charitable donations. We highlight the limitations of traditional donation systems and how a blockchain-based system can help overcome these challenges. The functionality of smart contracts in donation tracking, offering advantages such as automation, reduced transaction fees, and enhanced accountability, is elucidated. The decentralized and tamper-proof nature of blockchain technology is emphasized for increased transparency and fraud prevention. While elucidating the benefits, we also address challenges in implementing such a system, including the need for technical expertise and security considerations. By fostering trust and accountability, a donation-tracking system in Web 3.0, empowered by smart blockchain networks, aims to catalyze a profound positive impact in the realm of philanthropy.

## I. INTRODUCTION

Charitable giving has always been an integral part of human society, providing a means for individuals and organizations to support causes they care about and make a positive impact on the world [1]. Whether aiding vulnerable populations during crises, supporting scientific research, or fostering the arts, charitable giving has the potential to make a profound impact on lives and communities [2]. It is important to ensure that charitable giving is transparent, efficient, and trustworthy to maximize its impact and build public trust. The emergence of blockchain technology and smart contracts provides an opportunity to improve the traditional donation ecosystem, address its challenges, and create a more effective and equitable donation system [3–5].

In recent years, there has been growing concern about transparency and accountability in the charitable sector [6], as well as a need to improve the efficiency of the donation process. In response to these challenges, many organizations have begun exploring the use of technology to improve donation tracking and increase transparency [4,7].

Blockchain technology stands out as a promising solution in this domain. Operating as a decentralized digital ledger, blockchain ensures secure, transparent, and tamper-proof transactions. Through blockchain, donation tracking systems can establish an immutable record of all transactions, guaranteeing that donations reach their intended recipients. This heightened transparency cultivates trust within the charitable sector, potentially encouraging increased contributions to worthy causes [3,8].

Traditional donation systems have several limitations that can hinder transparency, accountability, and efficiency [9,10]. One of the primary challenges is the lack of transparency in the donation process. With traditional systems, it can be difficult for donors to track their contributions and ensure that they are being used for their intended purpose. This lack of transparency can also make it challenging for organizations to demonstrate the impact of their work and build trust with donors [11]. These systems also struggle with outdated processes, causing delays and restricting real-time access to information for donors. Security is another issue, as centralized platforms can be vulnerable to breaches and fraud. Moreover, these systems are slow to adapt, hindering quick responses to urgent needs. Another limitation of traditional donation systems is the high fees associated with processing transactions. Intermediaries such as banks, credit card companies, and payment processors can charge significant fees, reducing the amount of money that ultimately reaches the intended recipient. Additionally, the centralized nature of traditional platforms exposes them to single points of failure and vulnerability, making them susceptible to security breaches and fraud. These limitations include bureaucratic inefficiencies, security risks, inflexibility, and financial inefficiencies, hampering the overall performance of traditional donation platforms.

Blockchain technology provides a solution to these challenges by offering a more secure, transparent, and accountable donation tracking system [12,13]. In the context of donation tracking, the use of smart contracts also ensures that all transactions are transparent and immutable, meaning that once a transaction is recorded on the blockchain, it cannot be altered or deleted [14,15]. This provides an additional layer of security and accountability, enabling all parties involved to access transaction details and verify their accurate execution. Moreover, the decentralized nature of blockchain eliminates the need for intermediaries, reducing transaction fees and ensuring that a larger portion of the donation directly reaches its intended recipient [16]. Incorporating Web 3.0 principles further enhances the efficiency of donation systems by decentralizing data management, providing decentralized identifiers (DID), and fostering trust through incentive and consensus algorithms. Together, these technologies revolutionize traditional donation systems, addressing their limitations and creating a more streamlined, secure, and transparent process.

Another significant advantage is the reduction in transaction fees. Blockchain and smart contracts eliminate intermediaries, leading to substantial cost savings. This fee reduction directs more funds toward beneficiaries, amplifying the impact of donations. This reduction in fees means that more funds can go directly towards the intended beneficiaries, ultimately increasing the impact of the donation.

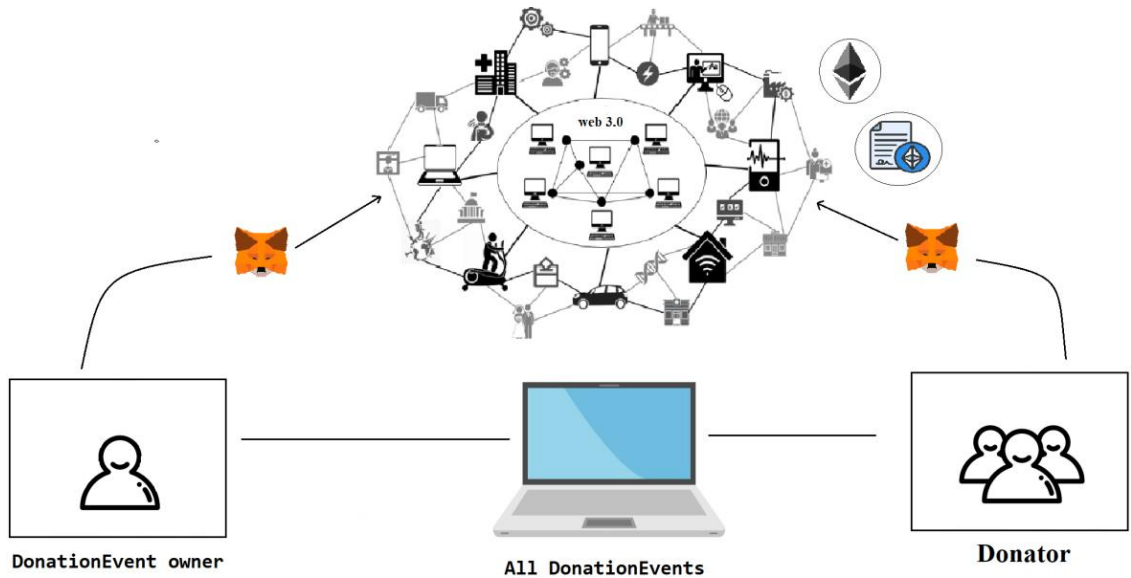

**Figure 1:** Blockchain based donation tracking system (DonateBlocks)

The proposed system (DonateBlocks) introduces a transparent, efficient, and secure system to track contributions, aiming to revolutionize traditional donation systems. By developing a new donation tracking system, we enhance efficiency and transparency in the process. As depicted in Figure 1, the system allows donors to track their donations in real time and provides regular updates on how their contributions are being utilized. Additionally, the system allows donors to connect their Metamask wallet to the platform, providing a secure and seamless way to access the Web 3.0 ecosystem. This system has the potential to revolutionize how donations are tracked and managed and could have a significant impact on the overall effectiveness of charitable organizations.

In this paper, we will explore the benefits of using blockchain technology, smart contracts, and Web 3.0 in donation-tracking systems. We will examine the current issues faced by traditional donation-tracking systems, such as lack of transparency and accountability, and explore how blockchain technology, smart contracts, and decentralized identities can address these issues. Additionally, we will discuss the benefits of smart contracts in automating donation transactions and ensuring fund utilization according to their intended purpose. Furthermore, we will provide a comprehensive overview of the technical aspects of implementing a blockchain-based donation tracking system, including the use of digital signatures and decentralized networks. The major contributions of this article are as follows:

- Providing a comprehensive overview of the use of blockchain technology and smart contracts, Web 3.0 principles, Metamask wallets, and DID in the donation tracking system.
- Analyzing the benefits of using blockchain technology and smart contracts in terms of transparency, security, and accountability in donation tracking
- Discussing the potential impact of blockchain technology, smart contracts, and emerging technologies like Web 3.0 and DID can have on the future of donation tracking and the nonprofit sector.

To facilitate understanding, this article has been organized into several sections. Section 2 provides a comprehensive review of related work, while Section 3 discusses the blockchain-based donation tracking system. Section 4 focuses on the implementation details of the proposed system architecture, including the system setup, smart contract, smart contract deployment, and possible vulnerabilities. In Section 5, we evaluate the performance of the system. Finally, the article concludes in Section 6.

## II. RELATED WORK

In recent years, there has been a growing interest in using blockchain technology for various applications [17], including donation-tracking systems for charitable giving. Several studies have explored the potential benefits and challenges of implementing blockchain-based donation tracking systems. One study found that a blockchain-based system can provide increased transparency, security, and efficiency in the donation process [18]. Another study suggested that blockchain can help address the issue of trust and accountability in charitable organizations [19].

Previous research has explored the potential applications of blockchain technology in the non-profit sector. One study conducted by the Charities Aid Foundation (CAF) found that blockchain technology could be used to create a more transparent and accountable donation system [20]. The study suggested that blockchain could provide a tamper-proof ledger for tracking donations, which would help build trust with donors and increase transparency in the donation process.

Furthermore, a report published by the United Nations Development Programme (UNDP) highlighted the potential of blockchain technology in promoting social and economic development [21]. The report suggested that blockchain could be used to create a more efficient and transparent donation system, which would enable a more effective allocation of resources to social causes.

Overall, previous research has shown that blockchain technology has the potential to revolutionize the way donations are tracked and managed in the non-profit sector. The use of blockchain technology can increase transparency, reduce transaction fees, and improve accountability, ultimately leading to a more efficient and effective donation ecosystem.

Similar donation-tracking systems that use blockchain technology and smart contracts have been developed by various organizations and companies. One such example is the BitGive Foundation [22], which has created a donation platform called GiveTrack that utilizes blockchain technology to provide transparency and accountability for charitable donations.

Another example is the Binance Charity Foundation [23], which has developed a blockchain-based donation platform called Binance Charity. The platform uses smart contracts to ensure that donations are used for their intended purpose and provides donors with real-time updates on the impact of their contributions.

Additionally, the United Nations World Food Programme (WFP) has implemented a blockchain-based donation platform called Building Blocks [24]. The platform uses smart contracts to track food deliveries and ensure that donations are being used to provide food assistance to refugees.

These examples demonstrate the potential of blockchain technology and smart contracts to create more transparent and efficient donation-tracking systems. By leveraging these technologies, organizations and individuals can have greater confidence in their charitable giving, knowing that their contributions are being used for their intended purpose.

There are several academic articles and research papers that explore the potential benefits and limitations of blockchain technology and smart contracts in the non-profit sector. One such paper is "Blockchain for Social Impact: Moving Beyond the Hype," which examines the potential for blockchain technology to create a more efficient and transparent non-profit sector [25]. The paper discusses the benefits of using blockchain technology, such as increased transparency, reduced transaction costs, and greater accountability, while also acknowledging the challenges that must be addressed, such as scalability and regulatory compliance.

Furthermore, [26] proposes a system called Charity-Chain, which is a decentralized network built on the Ethereum blockchain that aims to increase transparency and accountability in social organizations. The system uses smart contract-based incentives to ensure the impact of projects is independently verified and accessible to all parties involved, making it easier for funders to monitor their transactions and restore trust in social organizations.

Overall, these papers demonstrate the growing interest in using blockchain technology and smart contracts to improve the efficiency and accountability of the non-profit sector. They provide valuable insights into the potential benefits and challenges of implementing these technologies and can inform the development of future donation-tracking systems.

In addition to the aforementioned studies and initiatives, several research endeavors have explored innovative solutions to enhance blockchain systems' functionality and efficiency. One such initiative, VQL (Verifiable Query Layer), introduces efficient and verifiable cloud query services tailored for blockchain systems [27]. By deploying a middleware layer in the cloud, VQL facilitates both efficient data query services and ensures the authenticity of query results. Through cryptographic techniques and blockchain integration, VQL provides a secure and reliable means of accessing blockchain data, contributing to improved efficiency and trust in blockchain-based systems.

In a similar vein, vChain+ focuses on optimizing verifiable blockchain boolean range queries, enhancing the querying capabilities of blockchain systems [28]. By implementing innovative techniques for verifying boolean range queries on the blockchain, vChain+ enables more precise and reliable data retrieval, thereby improving the overall utility and effectiveness of blockchain-based applications.

Complementing these efforts, BlockShare presents a blockchain-empowered system designed for privacy-preserving and verifiable data sharing [29]. By leveraging blockchain technology, BlockShare ensures the privacy and integrity of shared data while enabling transparent and auditable data exchanges. This system contributes to enhancing data-sharing practices across diverse domains while maintaining confidentiality and trust among participating entities.

Collectively, these research efforts signify the ongoing exploration and development of advanced solutions to enhance blockchain systems' capabilities. Through initiatives like VQL, vChain+, and BlockShare, researchers and practitioners aim to address critical challenges and unlock new opportunities for leveraging blockchain technology in various domains. These endeavors underscore the dynamic nature of blockchain research and its potential to drive innovation and progress in diverse fields.

### III. PROPOSED SYSTEM: BLOCKCHAIN-BASED DONATION TRACKING SYSTEM

Our revolutionary platform, DonationBlocks, employs blockchain technology to redefine how donations are tracked, introducing principles of decentralization, transparency, immutability, and security into the donation process. This section addresses the challenges of traditional donation systems and illustrates how our proposed blockchain-based donation tracking system mitigates these issues. Traditional donation systems face challenges such as limited transparency, accountability, and high transaction fees. Donors often struggle to trace their contributions, hindering trust in the donation process. Transaction intermediaries amplify costs, reducing the overall impact of donations. DonationBlocks leverages blockchain's core attributes to address these challenges. The decentralized network serves as an incorruptible ledger, recording and verifying donation transactions in real time. Smart contracts automate processes, eliminating intermediaries, reducing fraud risks, and providing donors with a transparent view of their contributions.

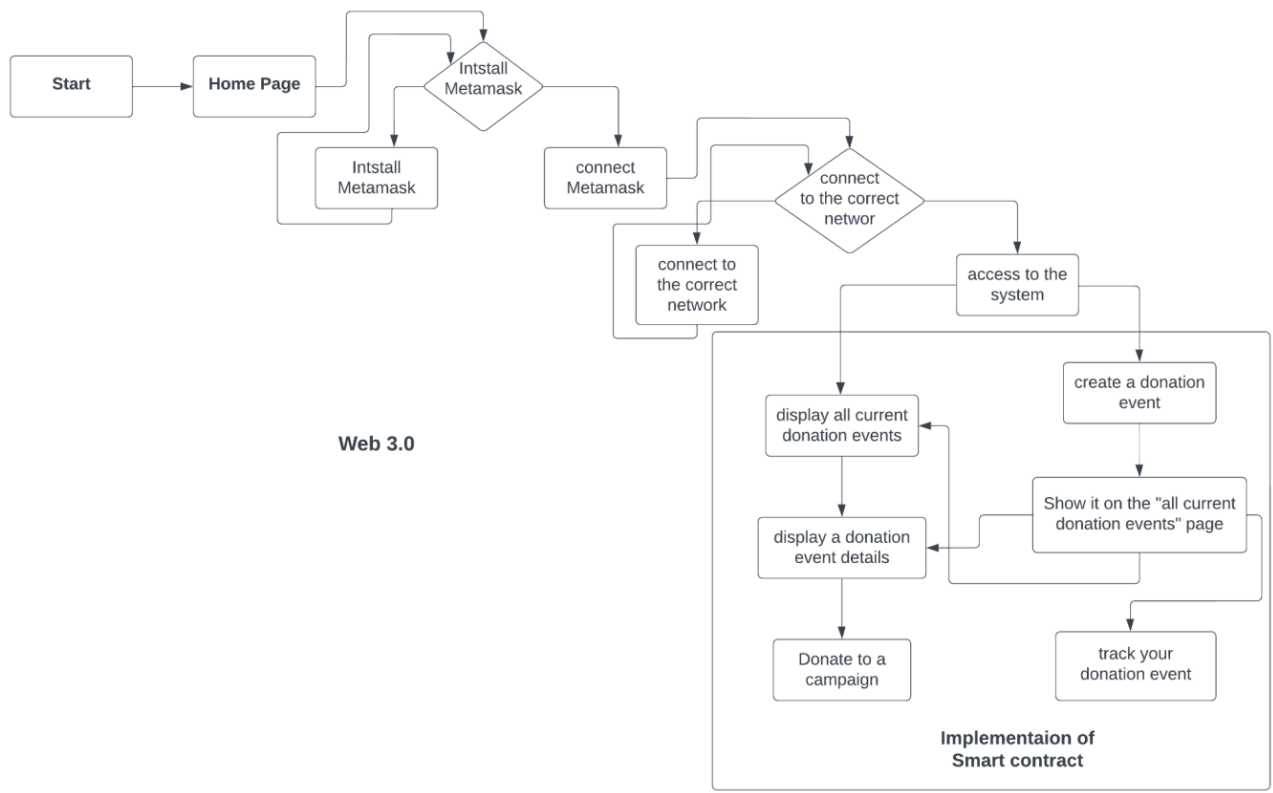

**Figure 2:** Donation tracking system workflow

As shown in Figure 2, the blockchain network is the foundation of the proposed blockchain-based donation tracking system, providing a decentralized platform for recording and verifying donation transactions. The network comprises a distributed ledger that securely stores all donation transaction data, with smart contracts regulating the rules and regulations governing the donation process to ensure transparency and eliminate the possibility of manipulation or fraud. The platform's workflow can be illustrated in Figure 3, and Figure 4, providing users with a clear and visual representation of the platform's features and functionalities at each step.

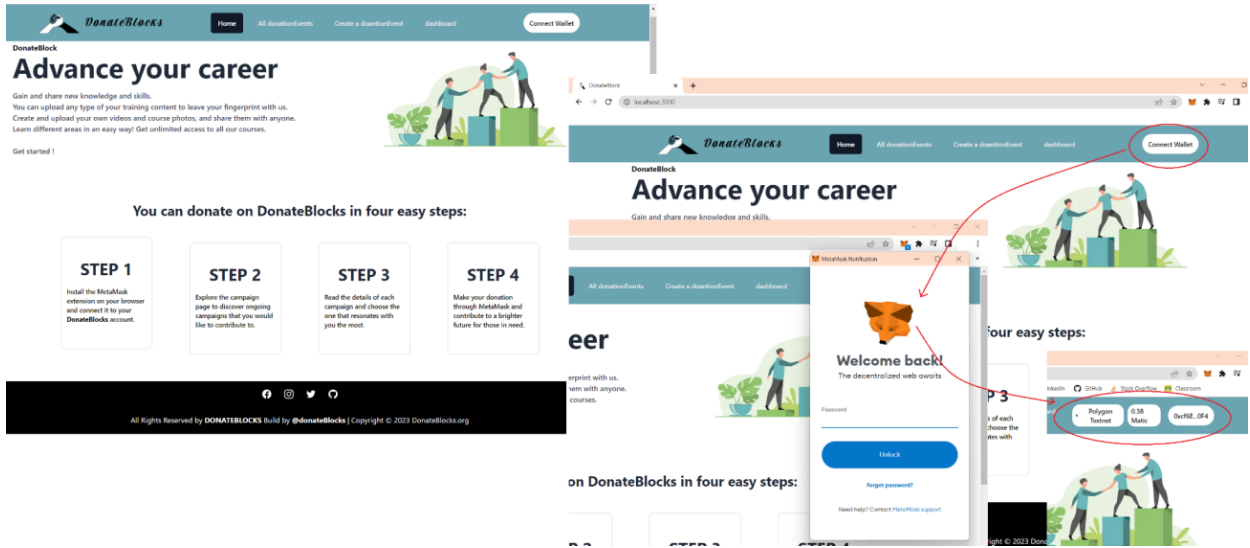

**Figure 3:** Main page of DonateBlocks platform

Upon visiting the main page of the DonateBlocks platform, as shown in Figure 3, users are greeted with a clear and intuitive interface that outlines the platform's workflow in a series of easy-to-follow steps. The page features a prominent navbar at the top of the screen, which provides users with quick access to key platform features. One of the main buttons in the navbar is the "Connect Wallet" button, which allows users to connect their digital wallet to the platform. Once a user has connected their wallet, the "Connect Wallet" button transforms to display important wallet information, including the wallet address, the amount of Matic cryptocurrency in the wallet, and the name of the network (in this case, the Polygon network). In addition to the "Connect Wallet" button, the navbar also includes four other buttons that take users to different pages within the platform. The "Home" button returns users to the main page of the platform, while the "All DonationEvents" button displays a list of all ongoing donation events on the platform. The "Create a DonationEvent" button allows users to create their donation event, while the "Dashboard" button takes users to a personalized dashboard that displays their ongoing donations and fundraising progress.

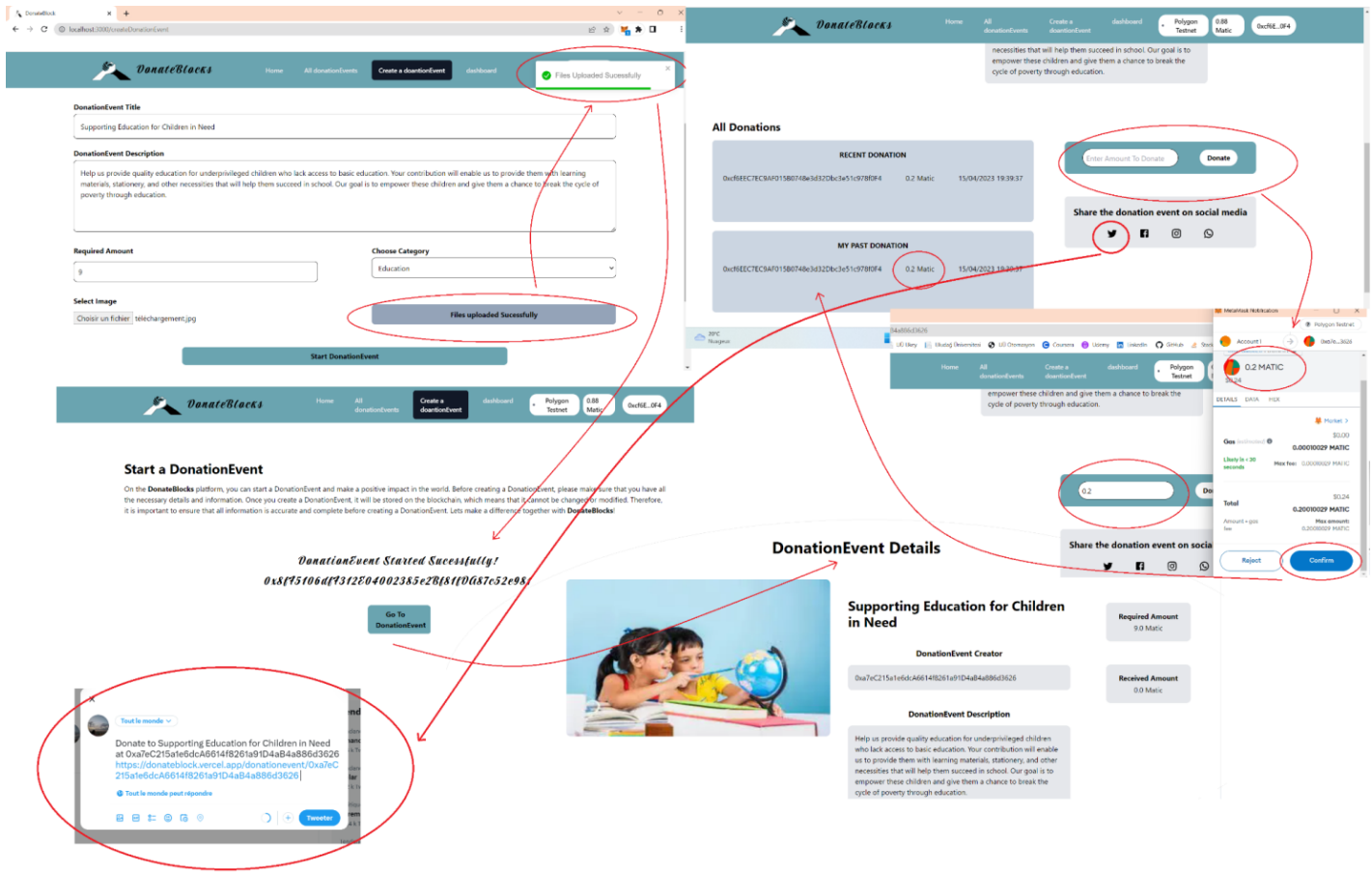

**Figure 4:** Creating and sharing a DonationEvent

Figure 4 provides a step-by-step guide for creating a donation event on the platform. Users can easily fill out the donation event title, description, required amount in Ether, and upload an image for their event. After filling out the form, the image is uploaded to IPFS and a donation event is started. Once the donation event is created, users are taken to a loading page where the transaction is executed. Once the transaction is confirmed, users are taken to the donation event details page. Here, users can see everything related to the donation event, including the current amount raised and the number of donors. To donate to the event, users can simply click on the "Donate" button and then use their MetaMask wallet to send Ether to the donation event's address. All donation transactions are shown on the page, and users can see the most recent transactions as well as their donation history to the event. Additionally, users can easily share the donation event on social media platforms like WhatsApp, Instagram, Twitter, and Facebook by using the IPFS protocol. In the figure, we demonstrate how to share the event on Twitter using IPFS.

## IV. IMPLEMENTATION DETAILS

The implementation of the DonationBlocks system involves a comprehensive use of various programming languages and frameworks to achieve a robust and functional donation tracking platform. The back-end of our system is implemented using Solidity, a programming language specifically designed for creating smart contracts on the Ethereum blockchain. These smart contracts govern various aspects of the system, including the creation and management of donation events, handling donations, and tracking donor information.

To develop, test, and deploy smart contracts, we utilize the Ethereum blockchain and the Polygon network. Ethereum serves as a popular platform for decentralized applications, while Polygon provides a layer-2 scaling solution that enhances transaction speed and cost efficiency. This combination of Ethereum and Polygon ensures a robust and secure infrastructure for our donation tracking system, enabling transparency, accountability, and efficiency in managing charitable donations. To facilitate the development and testing of our smart contracts, we use Hardhat, a development environment. Hardhat offers built-in functionality for compiling, deploying, and testing smart contracts and can be extended through plugins. With Hardhat, we can compile our Solidity code, deploy the contracts, and execute automated tests, ensuring the reliability and functionality of our system.

Our back-end implementation, powered by Solidity, Ethereum, Polygon, and Hardhat, forms a strong foundation for effective management of donation events and the integrity of the donation tracking system. For seamless communication between the front-end and back-end components, we adopt Web3.js, a JavaScript library providing a user-friendly interface for interacting with the Ethereum blockchain. Various Web3 libraries aid in testing and deploying smart contracts on the Ethereum blockchain, ensuring smooth integration between different system components. Ethereum-waffle, a library designed for testing smart contracts, simplifies testing efforts, while Ethers, a JavaScript library, facilitates front-end code interactions with smart contracts, enabling users to seamlessly donate and view events within the system. Leveraging these Web3 libraries ensures efficient development and testing, resulting in a robust and user-friendly donation tracking system on the Ethereum blockchain.

Incorporating IPFS (InterPlanetary File System) for decentralized file storage, our donation tracking system allows users to store and retrieve images and media files securely. Each uploaded file is assigned a unique hash stored on the blockchain, ensuring accessibility even if the original uploader or storage node is unavailable. This integration enhances the system's functionality with reliable and secure storage.

Additionally, we integrated social media platforms into our system, enabling users to share their donations and promote donation events. Leveraging IPFS's social media sharing feature, users can easily share content from the platform on platforms like Facebook, Twitter, WhatsApp, and

Instagram. This integration aims to increase visibility and encourage wider participation in donation events. The integration with social media was a smooth process, providing an effective tool to maximize the impact of hosted donation events.

Furthermore, our system integrates Decentralized Identifiers (DID) for user registration and authentication, empowering users with control over their identity. This registration process is seamlessly aligned with the Metamask wallet, ensuring a secure and user-friendly experience. Social media platforms are also integrated, enabling users to share their donations and promote events, maximizing the impact of charitable activities in the digital space. The implementation incorporates Web3 technologies, fostering a decentralized and trustless environment, aligning with the ethos of the evolving Web 3.0 paradigm.

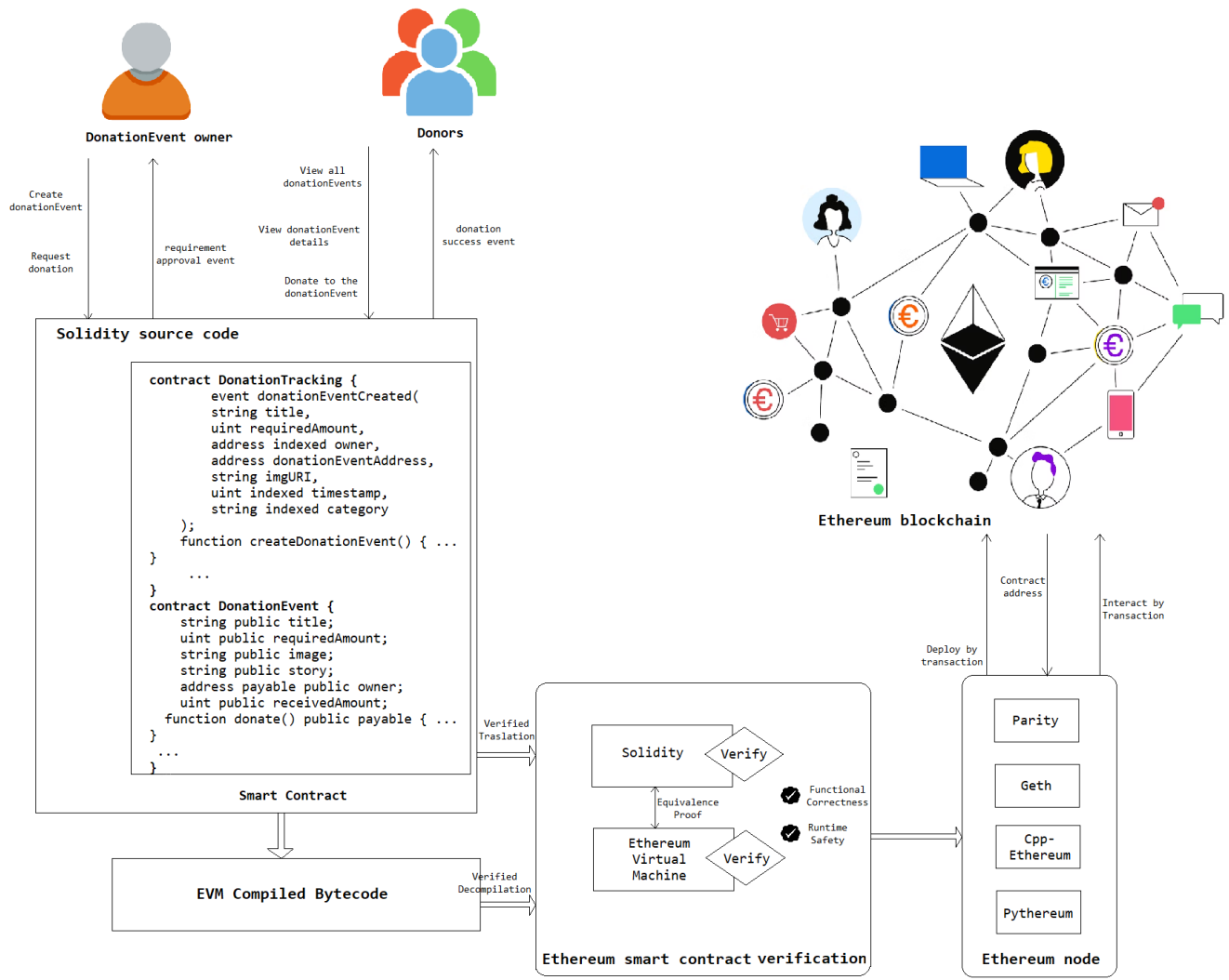

**Figure 5:** "DonateBlocks" system architecture

The system architecture of the "DonateBlocks" platform is illustrated in Figure 5. It enables donation event owners to create new events by deploying smart contracts to the Ethereum blockchain. These smart contracts store relevant information such as the event title, required amount, image, story, and total donations received. Donors interact with the smart contract by donating Ether to the contract address of the donation event.

Within proposed system, two distinct types of smart contracts are employed: DonationEvent contracts and DonationTracking contracts. DonationEvent contracts are created by donors to govern the donation process, specifying the donation amount, recipient organization, and any additional instructions or conditions. These contracts execute automatically without intermediaries. DonationTracking contracts are created by recipient organizations and provide a public record of all donations received. They ensure transparency and accountability, allowing donors to view their donation history and facilitating accurate financial reporting by recipient organizations. These contracts include parameters such as a minimum threshold and deadline for a successful donation campaign.

The Solidity source code for DonationTracking and DonationEvent contracts can be compiled into EVM bytecode using a Solidity compiler. The bytecode is then deployed to the Ethereum blockchain using an Ethereum node through a transaction to the contract address. Contract deployment can be verified on the Ethereum blockchain by confirming bytecode-source code correspondence.

In summary, the strength of the "DonateBlocks" platform lies in its adept utilization of cutting-edge technologies, establishing a robust and efficient donation tracking system. Ethereum, as the chosen blockchain platform, provides the decentralized infrastructure necessary for seamless and scalable operations. The implementation of Solidity, a programming language designed for smart contracts, empowers the system with the capability to govern various aspects, from creating and managing donation events to tracking donor information. The integration of Web3 libraries facilitates smooth communication between the front-end and back-end components, ensuring a user-friendly interface for interactions with the Ethereum blockchain. Furthermore, the incorporation of Decentralized Identifiers (DIDs) enhances the platform's security and privacy features, allowing users to register and control their identities securely. This amalgamation of Ethereum, smart contracts, Web3, and DIDs collectively contributes to the robust, transparent, and secure nature of the "DonateBlocks" donation tracking system.

## V. PERFORMANCE EVALUATION

In this section, we will compare the performance of our system with other similar systems benchmarks. It's important to note that many factors can affect the performance of a blockchain-based system, such as the size of the network, the complexity of the smart contracts, and the hardware specifications of the nodes.

Table 1 provides a comparison between the existing offline and online donation systems and the proposed DonateBlocks system based on several key factors. By comparing the systems based on these factors, we can gain insights into the strengths and weaknesses of each system and understand how the proposed DonateBlocks system can potentially improve upon the existing systems.

**Table 1:** Comparison of donation management technologies

| Division | Existing offline system | Existing online system | "DonateBlocks" proposed system |
|---|---|---|---|
| **User-friendliness** | Limited user-friendliness due to manual processes and lack of user interface | Generally user-friendly, but may require technical knowledge to navigate and use effectively | User-friendly interface with clear instructions and easy-to-use features, resulting in a positive user experience |
| **Cost-effectiveness** | Higher costs associated with manual processes, such as physical record-keeping and paperwork | Varies depending on the platform used and the level of customization required | Higher than the existing online system, as it eliminates the need for intermediaries and reduces transaction fees |
| **Customizability** | Limited customizability, not easily adapted to other organizations' requirements | Moderate customizability, limited in scope, or may require additional fees | High customizability, modular design easily configurable to meet different organizations' needs |
| **Flexibility** | Lack of flexibility, manual processes for changes or updates | Some flexibility, but limitations due to technology or vendor-specific solutions | Highly flexible and customizable, accommodates changes and updates easily |
| **Scalability** | Significant challenges, manual processes, and resource-intensive | More scalable, but limitations due to technology or architecture | Designed to be highly scalable with a decentralized architecture that handles increased data and users effectively |
| **Reliability** | Prone to human error, maintenance and upgrades can be costly | Can vary, downtime and technical issues can occur | Inherently reliable with secure blockchain transactions, decentralized nature reduces downtime and technical issues |

By considering these factors, it becomes apparent that the DonateBlocks proposed system offers significant improvements over the existing offline and online systems. It combines user-friendliness, cost-effectiveness, customizability, flexibility, scalability, and reliability to create a robust and efficient donation-tracking system. The integration of blockchain technology enhances transparency, privacy, and security, making it a promising solution for organizations seeking a more efficient and trustworthy way to manage donations.

As shown in the Figure 6, the donation event owner creates a new DonationEvent by calling the createDonationEvent function in the smart contract. The function takes as input the owner's address and name, title, description, target amount, deadline, and image, and creates a new instance of the DonationEvent struct. The MetaMask wallet provides secure storage and management of cryptocurrencies and enables donors to register and make donations to their chosen charitable cause. The donor first connects their MetaMask wallet to the platform, and then selects the donation event they wish to donate to. The smart contract manages the creation and tracking of donation events and donations. It includes functions such as createDonationEvent, donateToDonationEvent, getDonors, and getDonationEvents. When a donor donates by calling the donateToDonationEvent function with the required parameters, the smart contract updates the DonationEvent struct with the new donation and the corresponding donor's address. The smart contract then updates the DonationEvent struct with the new donation and the corresponding donor's address, ensuring secure tracking and recording of all donations on the Ethereum blockchain.

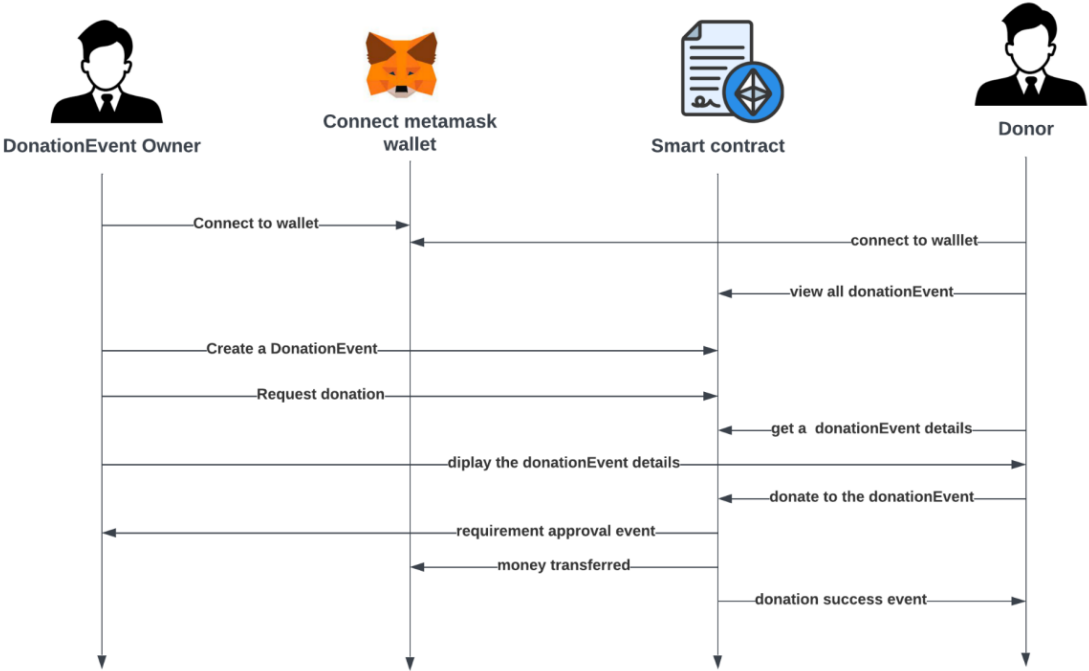

**Figure 6:** Blockchain-based donation system (DonateBlocks) sequential diagram

The DonateBlocks platform offers a secure and decentralized solution for tracking donations, leveraging blockchain technology and smart contracts. It provides transparency, trust, and accountability in the donation process. With features such as MetaMask integration, user-friendly interface, donation tracking and reporting, payment gateway integration, and smart contract management, the platform ensures secure storage and management of funds, real-time updates on campaign progress, and convenient donation methods. By optimizing smart contract code and using a high-performance blockchain network, the platform is scalable, cost-effective, and secure. Overall, DonateBlocks has the potential to enhance transparency, accountability, and trust in donations, ensuring they are used for their intended purposes.

To evaluate the performance of the proposed architecture, various analyses were conducted. These analyses focused on the transaction latency and throughput metrics, which are commonly used in the literature. Figure 7 presents the transaction latency results for the proposed architecture. As expected, the results indicate that latency increases with the number of transactions in the system. In Blockchain-based donation system, increasing the number of users and transactions requires more node approval time and transaction submission time. In our system, we examined the number of transactions between 5 and 50 and observed latency variations between approximately 88 seconds and 180 seconds.

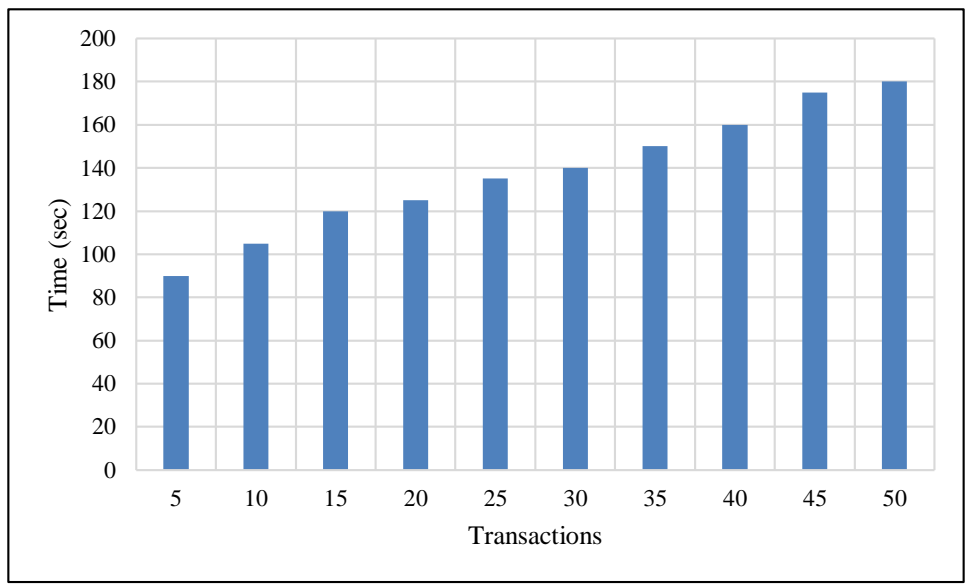

**Figure 7:** Transaction latency of the Blockchain-based donation system

Figure 8 shows the throughput results of the proposed architecture based on the number of transactions. Throughput is calculated in terms of Transactions per Minute (TPM). The results indicate that the Blockchain-based donation system requires 6 TPM for 5 transactions. It was observed that the TPM values also increase as the number of transactions increases. Approximately 40 TPM is required for 50 transactions. The obtained results are consistent with the studies in the literature [30,31]. These results demonstrate that the proposed architecture may lead to higher transaction volumes and delays compared to traditional donation systems. However, it is also known that the system will offer significant contributions considering the advantages given in Table 1.

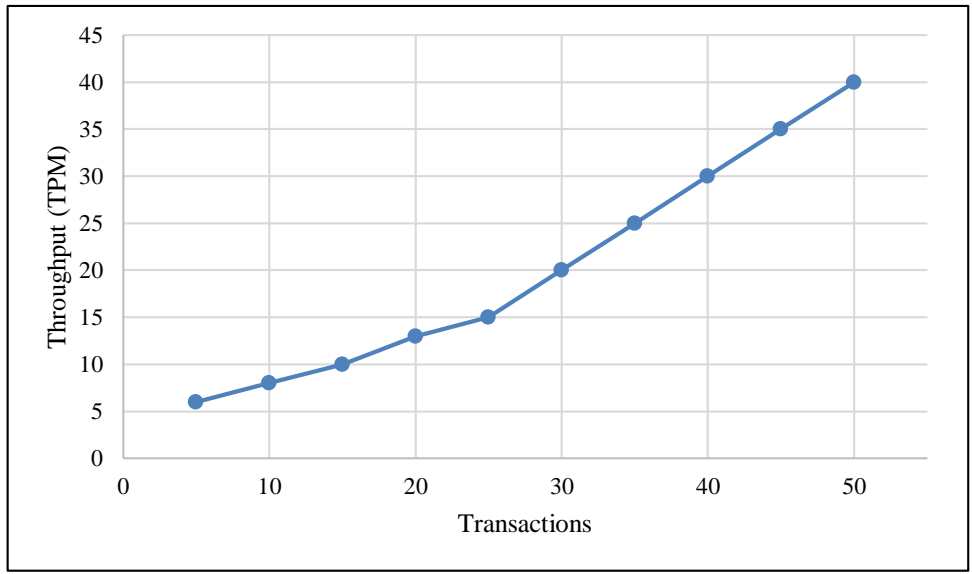

**Figure 8:** Throughput of the Blockchain-based donation system

## VI. CONCLUSION

In conclusion, a donation-tracking system using smart contracts and blockchain can provide a transparent, secure, and decentralized solution for tracking charitable donations [26]. By leveraging the benefits of blockchain technology and smart contracts, the system can increase transparency, accountability, and trust in the donation process [32]. The proposed system architecture involves several components, including a donor interface, smart contracts, a blockchain network, nodes, a recipient organization interface, and a payment gateway. By working together, these components provide a tamper-proof and decentralized ledger for tracking donations.

The performance of a donation tracking system using smart contracts and blockchain can be evaluated based on factors such as transaction speed, transaction fees, scalability, and security. By optimizing smart contract code, using a high-performance blockchain network, and automating many of the processes involved in donation tracking, the system can provide a scalable, low-cost, and secure solution for tracking donations. Overall, a donation-tracking system using smart contracts and blockchain has the potential to increase transparency, accountability, and trust in the donation process, ultimately helping to ensure that donations are used for their intended purposes. As such, it represents an important step forward in the field of charitable giving and has the potential to make a significant impact in the world of philanthropy.

In addition to the current implementation of the donation-tracking system using smart contracts and blockchain, there are opportunities for future enhancements and optimizations, particularly in leveraging layer 2 solutions on the Ethereum blockchain. Integrating the system with layer 2 scaling solutions, such as Ethereum's Optimistic Rollups or zk-Rollups, can significantly improve transaction speed and reduce transaction fees. By offloading transaction processing to layer 2 networks, the system can achieve near-instantaneous transaction confirmation and substantially lower costs, making it more accessible and efficient for users. Furthermore, exploring interoperability with other blockchain networks and decentralized finance (DeFi) protocols could expand the system's capabilities and interoperability, enabling seamless integration with a broader ecosystem of financial services and applications. Additionally, ongoing research and development efforts in blockchain technology may introduce new features and optimizations that could further enhance the performance and functionality of the donation-tracking system. Therefore, future work could focus on integrating these advancements into the system to continuously improve its scalability, usability, and overall effectiveness in facilitating charitable donations.

**CRediT authorship contribution statement**
Chaimaa Nairi: Conceptualization, Methodology, Software, Writing – original draft, Visualization, Writing – review & editing. Murtaza Cicioğlu: Conceptualization, Methodology, Software, Writing – original draft, Visualization, Writing – review & editing. Ali Çalhan: Conceptualization, Methodology, Software, Writing – original draft, Visualization, Writing – review & editing.

**Declaration of Competing Interest**
The authors declare that they have no known competing financial interests or personal relationships that could have appeared to influence the work reported in this paper.

**Data Availability Statement**: Data sharing not applicable – no new data generated

**Funding Statement:** This research received no specific grant from any funding agency in the public, commercial, or not-for-profit sectors.

## References

[1] TransparentHands, Concept, Purpose, and Importance of Charity in Our Society, (2018). https://www.transparenthands.org/concept-purpose-and-importance-of-charity-in-our-society/ (accessed February 19, 2023).
[2] P. Vallely, How philanthropy benefits the super-rich, Guard. (2020). https://www.theguardian.com/society/2020/sep/08/how-philanthropy-benefits-the-super-rich (accessed February 19, 2023).
[3] C. Ugaz-Burga, R. Valverde-Grados, D. Cardenas-Salas, Blockchain and smart contract for donation traceability, in: Proc. 5th Int. Conf. Trends Electron. Informatics, ICOEI 2021,

2021. https://doi.org/10.1109/ICOEI51242.2021.9453016.

[4] J. Sung, G.W. Bock, H.M. Kim, Effect of blockchain-based donation system on trustworthiness of NPOs, Inf. Manag. 60 (2023). https://doi.org/10.1016/j.im.2023.103812.

[5] A. Seo, Y. Son, Y. Lee, J. Jeong, A Blockchain Enabled Personal Donation System Development Scheme, J. Syst. Manag. Sci. 12 (2022). https://doi.org/10.33168/JSMS.2022.0108.

[6] A. Singh, R. Rajak, H. Mistry, P. Raut, Aid, Charity and Donation Tracking System Using Blockchain, in: Proc. 4th Int. Conf. Trends Electron. Informatics, ICOEI 2020, 2020. https://doi.org/10.1109/ICOEI48184.2020.9143001.

[7] H. Saleh, S. Avdoshin, A. Dzhonov, Platform for tracking donations of charitable foundations based on blockchain technology, in: Proc. - 2019 Actual Probl. Syst. Softw. Eng. APSSE 2019, 2019. https://doi.org/10.1109/APSSE47353.2019.00031.

[8] E. Shaheen, M.A. Hamed, W. Zaghloul, E. Al Mostafa, A. El Sharkawy, A. Mahmoud, A. Labeb, M.O. Al Enany, G. Attiya, A track donation system using blockchain, in: ICEEM 2021 - 2nd IEEE Int. Conf. Electron. Eng., 2021. https://doi.org/10.1109/ICEEM52022.2021.9480649.

[9] Snowballfundraising, Donation Management, (2022). https://snowballfundraising.com/features-overview/#reporting (accessed March 20, 2023).

[10] J. Wan, Y. Lu, B. Wang, L. Zhao, How attachment influences users' willingness to donate to content creators in social media: A socio-technical systems perspective, Inf. Manag. 54 (2017). https://doi.org/10.1016/j.im.2016.12.007.

[11] B.D. Friedman, A.M. Wolcott, Secrecy and transparency in nonprofit organizations: if a nonprofit prefers secrecy, what does it want to hide?, 2018. https://core.ac.uk/download/pdf/228457672.pdf.

[12] A. Almaghrabi, A. Alhogail, Blockchain-based donations traceability framework, J. King Saud Univ. - Comput. Inf. Sci. 34 (2022). https://doi.org/10.1016/j.jksuci.2022.09.021.

[13] J. Karthika, S. Keerthana, A. Shali, Blockchain based Transparent Donating System, J. Inf. Technol. Digit. World. 5 (2023). https://doi.org/10.36548/jitdw.2023.2.003.

[14] M. Li, Y. Chen, L. Zhu, Z. Zhang, J. Ni, C. Lal, M. Conti, Astraea: Anonymous and Secure Auditing Based on Private Smart Contracts for Donation Systems, IEEE Trans. Dependable Secur. Comput. 20 (2023). https://doi.org/10.1109/TDSC.2022.3204287.

[15] V. Hiremath, Decentralised Application on Charity Using Blockchain, Int. J. Res. Appl. Sci. Eng. Technol. 11 (2023). https://doi.org/10.22214/ijraset.2023.51648.

[16] C.W. Cai, Disruption of financial intermediation by FinTech: a review on crowdfunding and blockchain, Account. Financ. 58 (2018). https://doi.org/10.1111/acfi.12405.

[17] B.M. Ramageri, M. V Arjunwadkar, Applications of Blockchain Technology in Various Sectors:A Review, Int. J. Futur. Gener. Commun. Netw. 13 (2020).

[18] J. Lee, A. Seo, Y. Kim, J. Jeong, Blockchain-based one-off address system to guarantee transparency and privacy for a sustainable donation environment, Sustain. 10 (2018). https://doi.org/10.3390/su10124422.

[19] A. Christie, Can Distributed Ledger Technologies Promote Trust for Charities? A Literature Review, Front. Blockchain. 3 (2020). https://doi.org/10.3389/fbloc.2020.00031.

[20] R. Davies, Submission to Treasury Select Committee Call for Evidence on Digital Currencies, Charities Aid Found. (2018). https://committees.parliament.uk/writtenevidence/89443/html/ (accessed April 25, 2023).

[21] C. Freire, Harnessing Blockchain for Sustainable Development: Prospects and Challenges,

in: Dev. United Nations Conf. Trade, United Nations Conference on Trade and Development, Geneva, 2021. https://unctad.org/system/files/official-document/dtlstict2021d3_en.pdf.

[22] C. Gallippi, BitGive, Bitgivefoundation. (2023). https://www.bitgivefoundation.org/ (accessed May 20, 2023).

[23] BINANCE CHARITY, Binance.Charity. (2023). https://www.binance.charity/ (accessed May 20, 2023).

[24] Building Blocks, United Nations World Food Program. (2023). https://innovation.wfp.org/project/building-blocks (accessed June 20, 2023).

[25] D. Galen, L. Boucherle, R. Davis, N. Do, B. El-Baz, K. Wharton, J. Lee, Blockchain for Social Impact - Moving Beyond the Hype, Stanford Grad. Sch. Bus. Cent. Soc. Innov. RippleWorks. (2018).

[26] N.S. Sirisha, T. Agarwal, R. Monde, R. Yadav, R. Hande, Proposed Solution for Trackable Donations using Blockchain, in: 2019 Int. Conf. Nascent Technol. Eng. ICNTE 2019 - Proc., 2019. https://doi.org/10.1109/ICNTE44896.2019.8946019.

[27] H. Wu, Z. Peng, S. Guo, Y. Yang, B. Xiao, VQL: Efficient and Verifiable Cloud Query Services for Blockchain Systems, IEEE Trans. Parallel Distrib. Syst. 33 (2022). https://doi.org/10.1109/TPDS.2021.3113873.

[28] H. Wang, C. Xu, C. Zhang, J. Xu, Z. Peng, J. Pei, vChain+: Optimizing Verifiable Blockchain Boolean Range Queries, in: Proc. - Int. Conf. Data Eng., 2022. https://doi.org/10.1109/ICDE53745.2022.00190.

[29] Z. Peng, J. Xu, H. Hu, L. Chen, BlockShare: A Blockchain Empowered System for Privacy-Preserving Verifiable Data Sharing, IEEE Data Eng. Bull. 45 (2022) 14–24. http://sites.computer.org/debull/A22june/p14.pdf.

[30] P. Sharma, R. Jindal, M.D. Borah, Blockchain-based distributed application for multimedia system using Hyperledger Fabric, Multimed. Tools Appl. 83 (2024) 2473–2499. https://doi.org/10.1007/s11042-023-15690-6.

[31] T.T.A. Dinh, J. Wang, G. Chen, R. Liu, B.C. Ooi, K.-L. Tan, BLOCKBENCH, in: Proc. 2017 ACM Int. Conf. Manag. Data, ACM, New York, NY, USA, 2017: pp. 1085–1100. https://doi.org/10.1145/3035918.3064033.

[32] S. Tunçer, A. Özdede, C. Karakuzu, Transparent Donation Management with Smart Contract-Based Blockchain, J. Eng. Res. Technol. 3 (2022).